\documentclass[12pt,a4paper]{article}

\pdfoutput=1
\usepackage[
top    = 2.50cm,
bottom = 2.50cm,
left   = 2.50cm,
right  = 2.50cm]{geometry}
\usepackage{graphicx}
\usepackage{tabularx}

\usepackage{bbm}
\usepackage{bbold}
\usepackage{latexsym}
\usepackage{amsmath}
\usepackage{amssymb}
\usepackage{cite}
\usepackage{breqn}
\usepackage[font={small,it}]{caption}
\usepackage{amsfonts}
\usepackage{xcolor}

\newcommand{\bra}[1]{\langle #1|}
\newcommand{\ket}[1]{|#1\rangle}

\newcommand{\eq}[2][ ]{\begin{equation}\label{#1}{\begin{split}#2\end{split}}\end{equation}}
\newcommand{\eql}[2]{\begin{equation}\label{#1}{\begin{split}#2\end{split}}\end{equation}}
\newcommand{\heading}[1]{\begin{center} \Large {#1} \end{center}}
\newcommand{\mo}{\mathcal{O}}
\newcommand{\hb}{\bar{h}}
\newcommand{\Tb}{\bar{T}}
\newcommand{\wb}{\bar{w}}
\newcommand{\zb}{\bar{z}}
\newcommand{\onov}[1]{\frac{1}{#1}}
\newcommand{\Sb}{\bar{S}}
\newcommand{\fb}{\bar{f}}
\usepackage{multirow}
\usepackage{pdfpages}

\usepackage{hyperref}

\makeatletter
\newcommand*\rel@kern[1]{\kern#1\dimexpr\macc@kerna}
\newcommand*\widebar[1]{%
  \begingroup
  \def\mathaccent##1##2{%
    \rel@kern{0.8}%
    \overline{\rel@kern{-0.8}\macc@nucleus\rel@kern{0.2}}%
    \rel@kern{-0.2}%
  }%
  \macc@depth\@ne
  \let\math@bgroup\@empty \let\math@egroup\macc@set@skewchar
  \mathsurround\z@ \frozen@everymath{\mathgroup\macc@group\relax}%
  \macc@set@skewchar\relax
  \let\mathaccentV\macc@nested@a
  \macc@nested@a\relax111{#1}%
  \endgroup
}

\def\bar{\widebar}

\begin{document}

\begin{titlepage}
  \vspace*{\fill}

\heading{\bf Entanglement Entropy and Conformal Collider Physics in 2D CFTs}

\vskip 1.4cm

\centerline{\it Talya Vaknin }
\bigskip
\centerline{ Weizmann Institute of Science, Rehovot
76100, Israel}

\vskip 4pt

\vskip 1.5cm

\begin{abstract}
We consider a conformal field theory in two dimensions in which an external perturbation is placed. We study the energy flux and entanglement entropy for one, two and multiple intervals and give a suggestion relating the two in some cases. We show that both the energy flux and the entanglement entropy exhibit a light-cone singularity which strongly depends on the regularization we choose.

\end{abstract}

  \vspace*{\fill}

\end{titlepage}

\pagebreak

\tableofcontents

\section {Introduction}
Almost a decade ago, Hofman and Maldacena \cite{Hofman:2008ar} came out with the novel idea of conformal collider physics. They considered an external perturbation of the theory localized in space and time and studied the properties of the state that was produced. Specifically, they calculated the energy flux at a distance from the excitation. In this paper we wish to ask what if we have access only to part of the data of this energy flux. In other words, if we can measure only in some subsystem, can we know anything about the entire system? This question is elegantly packed in the notion of entanglement entropy (EE) for a quantum state.
In this paper we will be interested in studying the EE in the existence of such a perturbation and suggest a connection between it and the energy flux.

Entanglement entropy has been widely studied in previous years mainly since it characterizes the quantum properties of a given system in a robust way. It is defined as the Von-Neumann entropy of the reduced density matrix $S_A=-Tr[\rho_A \log\rho_A]$. In the case of a conformal field theory in $1+1$ dimensions in its ground state with a subsystem $A$ of one interval of length $\ell$, the EE results in the famous equation\cite{Calabrese:2004eu} 
	\eq{S_A=\frac{c}{3}\log\frac{\ell}{a}+c'_1\quad,} that contains the universal signature $c$, the central charge of the conformal field theory. Another quantity of interest is the $n$-th R\'enyi entropy which is defined by $S_A^{(n)}=\log Tr[\rho_A^n]/(1-n)$ and coincides with the von-Neumann entropy when taking the limit $n\rightarrow 1$ when it is well defined. Unlike the single interval case, for a subsystem of multiple disjoint intervals, the EE depends, in general, on the full operator content of the theory. The two interval case was studied in \cite{Calabrese:2009ez,Calabrese:2010he, Headrick:2010zt} and a universal formula for $N$ disjoint intervals in the large $c$ limit was derived in \cite{Hartman:2013mia}.  

In the case motivated above, we wish to explore the behaviour of entanglement entropy for excited states. For a single interval, this has been formulated for locally excited states in \cite{Nozaki:2014hna}. Further calculations for states excited by scalar operators were done in \cite{Nozaki:2014uaa,Shiba:2014uia} and a remarkable relation has been found between the R\'enyi entropy for an excited primary state and its quantum dimension \cite{He:2014mwa} in rational CFTs. First attempts to extend these computations to descendant fields were made in \cite{Chen:2015usa,Caputa:2015tua} and a holographic computations were done in \cite{Sheikh-Jabbari:2016znt, Chen:2016kyz}. We will use some of these results to generalize and find a formula for descendant operators in the case of multiple disjoint intervals.

In two dimensions an interesting phenomenon occurs that was not analysed in generality before. This is the light-cone behaviour of the entanglement entropy as we will explain in the following. In the presence of an operator insertion, we expect that the EE will stay unchanged as long as the operator insertion doesn't cross the light-cone of the subsystem. Once the light-cone is crossed, there might and might not be a change in the EE, depending on the operator that has been inserted. In both cases, we encounter a singularity in the EE when crossing the light-cone itself. In the case of the second R\'enyi entropy, this light-cone singularity is exactly the same one we find in conformal blocks when one of the operators intersects the light-cones of two others simultaneously \cite{Hartman:2015lfa,Komargodski:2012ek,Fitzpatrick:2012yx,Li:2015itl}. This behaviour is not well understood yet and we will not go into analysing the divergence characteristics. We do notice that the energy flux in two dimensions is singular on the light-cone as well, which points out the connection between the two quantities. 

In this paper we analyse the two quantities, energy flux and entanglement entropy, in detail and then suggest a connection between them. 

Our paper is organized as follows. In section \ref{CCP} we calculate the energy flux in the presence of different excitations. We then calculate the EE for the energy momentum tensor in section \ref{EEEM} and generalize this result for other descendants in section \ref{EEdes}. Finally we conjecture a relation between the energy flux and the EE in section \ref{fluxEE}.

\section{Conformal Collider Physics}\label{CCP}
In this section we work in the spirit of Hofman and Maldacena \cite{Hofman:2008ar} in 2D CFTs. We will use three complementary methods of computations. The one will be by using Euclidean formulation throughout and analytically continuing to Minkowski at the very end. This formulation captures the light-cone behaviour. The second way will be to represent a descendant field by a unitary transformation acting on a primary operator. In this way we can generalize the former result to any descendant but the light-cone singularity behaviour is absent. In the third method we will use the AdS/CFT duality for excited  states using the Ba\~nados geometries \cite{Banados:1998gg} and compare to the result of the second approach.

\begin{figure}
  \centering
      \includegraphics[width=0.6\textwidth]{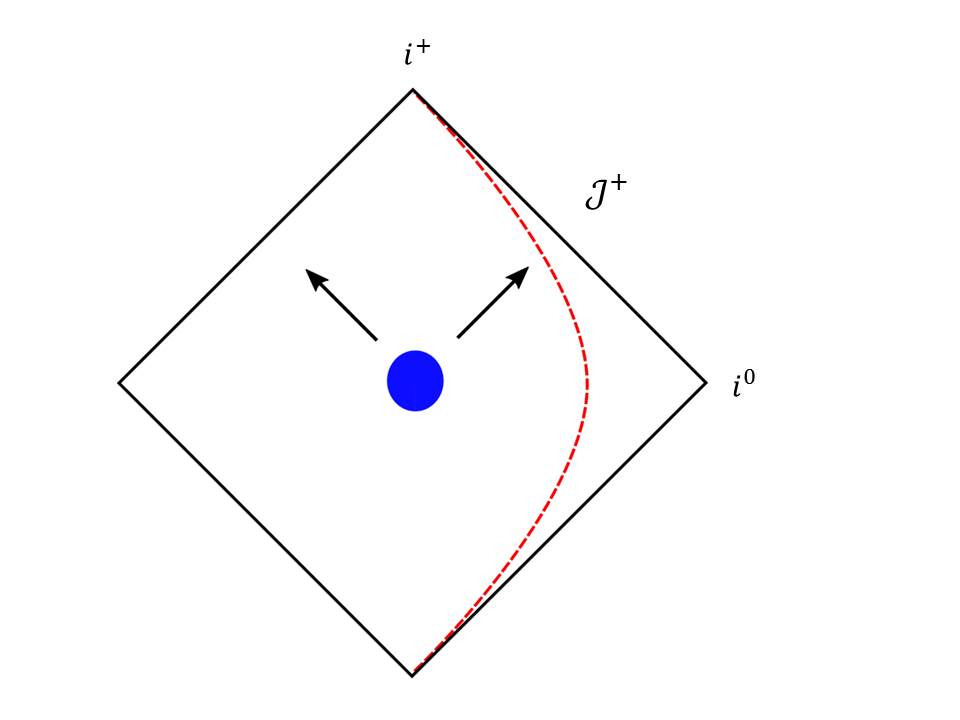}
  \caption{The Penrose diagram of Minkowski space in two dimensions. The red dotted line is $x=\ell$ where we place our calorimeter and the blue circle symbolizes the operator insertion we have at the origin. }
  \label{pen}
\end{figure}

We will place our "calorimeter" to measure the energy flux at some point in space $x=\ell$ over time $t$ and consider a perturbation coupled to an operator of the conformal field theory, localized at $x=0,\quad t=0$ (figure \ref{pen}.)

The energy flux computations in this case are three point functions- the operator insertions and the energy flux which is given by $T_{10}=-(T-\Tb)$. We will compute explicitly for an excitation created by an arbitrary primary field $\mo$ and by the energy momentum tensor $T$, a descendant of the identity field.

Since operators in Euclidean space aren't time ordered, we slightly continue time into the complex plane by an infinitesimal amount $\epsilon$. In this way we can impose the ordering that we are interested in by giving the operators later in time a bigger imaginary part. At the end of the calculations we will analytically continue by taking $\epsilon$ to zero.
Therefore our operators are placed at 
\eql{w's}{w_1=i\epsilon,\quad w_2=-i\epsilon\quad,\\
\wb_1=-i\epsilon,\quad \wb_2=i\epsilon\quad,
}
and our calorimeter at $(z,\zb)=(\ell-t,\ell+t)$. Notice that $z$ and $\zb$ are not complex conjugates of one another. This is due to the analytic continuation we perform from Euclidean space to Minkowski space.

For a primary field $\mo$ with conformal dimensions $(h,\hb)$, the energy flux is computed to be
\eql{flux_p}{-\frac{\langle\mo(w_1,\wb_1)\left(T(z)-\Tb(\zb)\right)\mo(w_2,\wb_2)\rangle}{\langle\mo(w_1,\wb_1)\mo(w_2,\wb_2)\rangle}&=-h\left(\frac{1}{z-w_1}-\frac{1}{z-w_2}\right)^2+\hb\left(\frac{1}{\zb-\wb_1}-\frac{1}{\zb-\wb_2}\right)^2\\
&=4h\left(\frac{\epsilon}{(\ell-t)^2+\epsilon^2}\right)^2-4\hb\left(\frac{\epsilon}{(\ell+t)^2+\epsilon^2}\right)^2\quad.}
We see that for any value of $t$ which is not on the light-cone, we can take the limit $\epsilon\rightarrow 0$ and the energy flux is zero. But for $t=\pm \ell$ we get a singularity in the energy flux. Due to this singularity the smearing of the operators into the complex plane is not well defined in the region close to the light-cone since the result will strongly depend on the choice of the smearing parameter.  

For an excitation created by the energy momentum field, the computation is straight forward. We notice that $\langle T(w_1)\Tb(\zb)T(w_2)\rangle=0$ therefore we have
\eql{em_flux}{-\frac{\langle T(w_1)T(z)T(w_2)\rangle}{\langle T(w_1)T(w_2)\rangle}=-2\frac{(w_1-w_2)^2}{(z-w_1)^2(z-w_2)^2}=8\left(\frac{\epsilon}{(\ell-t)^2+\epsilon^2}\right)^2\quad,}
which presents the same behaviour as a primary. 

Generalizing this result for any descendent is cumbersome. We will assume that for all descendants the energy flux preserves the same behaviour and obtains a singularity on the light-cone. At this point we will switch to a formalism that is valid only away from the light-cone since we cannot introduce the normalising parameter $\epsilon$.

We consider the coadjoint orbits which are a unitary representations of the Virasoro group \cite{Witten:1987ty, Balog:1997zz, Compere:2015knw, Sheikh-Jabbari:2016unm}. There is a one-to-one correspondence between these orbits and unitary Verma modules, therefore we will be able to identify a certain representative of the orbit with the highest weight state in the Verma module. We will be interested in those orbits that contain a constant representative, which corresponds to the primary operator in each family/orbit. States in the same orbit are related by a unitary transformation. More precisely, a descendant operator can be represented as a conformal transformation (unitary) operator $U_f$ acting on a primary state $|\mo\rangle$
\eq{|\mo_f\rangle=U_f|\mo\rangle\quad,}
where $f$ is the conformal map $z\rightarrow f(z)$, $\zb\rightarrow \bar{f}(\zb)$ associated with the unitary operator.
We can use the transformation of the energy momentum tensor
\eql{T_trans}{T(w)\rightarrow \widetilde{T}(w)=\left(\frac{dw}{dz}\right)^{-2}\left(T(z)-\frac{c}{12}\{w,z\}\right)\quad.}
to calculate the energy flux for a descendant of this type. Using normalized states, it is given by 
\eql{flux}{\langle\mo_f|& -\left(T(z)-\Tb(\zb)\right)|\mo_f\rangle=-\langle\mo|U^\dagger_f \left(T(z)-\Tb(\zb)\right)U_f|\mo\rangle=\\&-\left(\frac{\partial f}{\partial z}\right)^2\langle\mo|T(f(z))|\mo\rangle+\left(\frac{\partial \bar{f}}{\partial \zb}\right)^2\langle\mo|\Tb(\bar{f}(\zb))|\mo\rangle-\frac{c}{12}\left(\{f(z),z\}-\{\bar f(\zb),\zb\}\right)\quad,} 
where the last term is the Schwarzian derivative $\{f(z),z\}=\frac{f'''}{f'}-\frac{3}{2}\frac{(f'')^2}{(f')^2}$. We notice that in the case of a descendant of the vacuum, the energy flux reduced to the Schwarzian derivatives alone since the expectation value of the energy momentum in the vacuum vanishes. It simplifies further if the descendant is holomorphic or anti-holomorphic. If we consider an excitation created by the energy momentum tensor, the associated map is $f(z)=\onov{z}$ for which the flux vanishes in agreement with \eqref{em_flux} when taking the limit $\epsilon\rightarrow 0$ away from the light-cone. More generally, any $SL(2,\mathbb{C})$ transformation acting on the unitary operator will give a vanishing flux since the vacuum is invariant under these transformations and the Schwarzian derivative is zero.

Another way to approach this problem is to use the AdS/CFT duality. For generic excited states in 2d CFTs, the associated duals are the Ba\~nados geometries   \cite{Banados:1998gg}
\eql{banados}{ds^2=\ell_{AdS}^2\frac{dr^2}{r^2}-\left(rdx^+-\frac{\ell_{AdS}^2}{r} L_-(x^-)dx^-\right)\left(rdx^--\frac{\ell_{AdS}^2}{r}L_+(x^+)dx^+\right)\quad,}
where $x^\pm\in [0,2\pi]$ and $L_\pm=L_\pm(x^\pm)$ are smooth and periodic functions which encode the information about the geometry, which is locally $AdS_3$ with radius $\ell_{AdS}$. These geometries obey the Brown-Henneaux boundary conditions \cite{Brown:1986nw}, which imply that the central charge of the dual CFT is $c=\frac{3\ell_{AdS}}{2G}$. Therefore these semiclassical solutions of gravity corresponds to two dimensional CFTs with large central charge. Note that in order to compare to our two dimensional CFT we need to take $x^+$ to $z$ and $x_-$ to $\zb$ and ignore the periodicity. 
The correspondence between these geometries and the Virasoro coadjoint orbits has been studied in \cite{Sheikh-Jabbari:2016znt,Compere:2015knw,Sheikh-Jabbari:2016unm}. Specifically, the energy momentum tensors in the 2d CFT are related to the functions $L_-$ and $L_+$ by \footnote{In our definition we introduce an extra minus sign in order to be consistant with our choice of conventions.}
\eql{banadosT}{-\langle T(x^+)\rangle=\frac{c}{6}L_+\quad,\quad-\langle\bar{T}(x^-)\rangle=\frac{c}{6}L_-\quad.} 
A specific choice of $L_\pm$ will determine the corresponding primary state on the boundary. For a descendant excited state we perform a conformal transformation $h_\pm$ which obeys $h_\pm(x^\pm+2\pi)=h_\pm(x^\pm)+2\pi$ under which $L_\pm$ transform 
\eq{L_+(x^+)&\rightarrow h_+^{'2}L_+\left(h_+(x^+)\right)-\frac{1}{2}\{h_+(x^+),x^+\}\quad,\\L_-(x^-)&\rightarrow h_-^{'2}L_-\left(h_-(x^-)\right)-\frac{1}{2}\{h_-(x^-),x^-\}\quad,}
with the Schwarzian derivative defined as before. Plugging this into \eqref{banadosT} and identifying $f$ with $h_+$ and $\bar f$ with $h_-$ we get the same result we got using 2d CFT calculations \eqref{flux}.

Placing multiple calorimeters corresponds to  $N$-point correlation functions of $T-\Tb$.
\eq{\langle\mo_f|\left(T(z_1)-\Tb(\zb_1)\right)...\left(T(z_N)-\Tb(\zb_N)\right)|\mo_f\rangle\quad.}
For simplicity let's assume that the excited state is holomorphic. When performing the conformal transformation $f$, the energy flux takes the form
\eql{Nflux}{\left\langle\mo\left|\displaystyle\prod_{i=1}^{N}\left[\left(\frac{\partial f}{\partial z_i}\right)^2 T(f(z_i))+\frac{c}{12}\{f(z_i),z_i\}\right]\right|\mo\right\rangle\quad.}
For two calorimeters $N=2$ this includes the four point function
\eql{4p_}{\frac{\langle\mo(w_1)T(z_1)T(z_2)\mo(w_2)\rangle}{\langle\mo(w_1)\mo(w_2)\rangle}=
\frac{-8h\epsilon^2\left(z_1^2z_2^2-2h\epsilon^2(z_1-z_2)^2+\epsilon^2(z_1^2+z_2^2)+\epsilon^4\right)}{(z_1-z_2)^2(z_1^2+\epsilon^2)^2(z_2^2+\epsilon^2)^2}+\frac{c/2}{(z_1-z_2)^4}\quad,}
which has singularities on the two light-cones of the calorimeters and when they coincide. It also includes the three point function that we calculated before \eqref{flux_p} which has a contribution only on the light cone, and the two point function with no energy momentum insertions which contributes only the Schwarzian derivatives. Away from the singularities we get a non-zero contribution from the central charge term in \eqref{4p_} and from the Schwarzian derivative
\eql{2flux}{\frac{\langle\mo_f| T(z_1)T(z_2)|\mo_f\rangle}{\langle\mo_f\mo_f\rangle}=\left(\frac{\partial f}{\partial z_1}\right)^2\left(\frac{\partial f}{\partial z_2}\right)^2\frac{c/2}{(f(z_1)-f(z_2))^4}+\left(\frac{c}{12}\right)^2\{f(z_1),z_1\}\{f(z_2),z_2\}}
To summarize this part, we have found a diverging behaviour in the energy flux on the light-cone and a general formula away from it which agrees with the holographic computation. We have further generalized this result for multiple calorimeters.

\section{Entanglement Entropy for Energy Momentum Excited States}\label{EEEM}
We begin by calculating in detail the second R\'enyi entropy for an excited state, created by the energy momentum tensor. We will do this in a manner that is sensitive to the light-cone limit. This has been done to some extent in \cite{Chen:2015usa,Caputa:2015tua}, but the light-cone limit was not analysed. We follow the formulation and notations of \cite{He:2014mwa} but notice that their computation is done for rational CFTs while here it can be applied to any CFT. 

\begin{figure}
  \centering
      \includegraphics[width=0.8\textwidth]{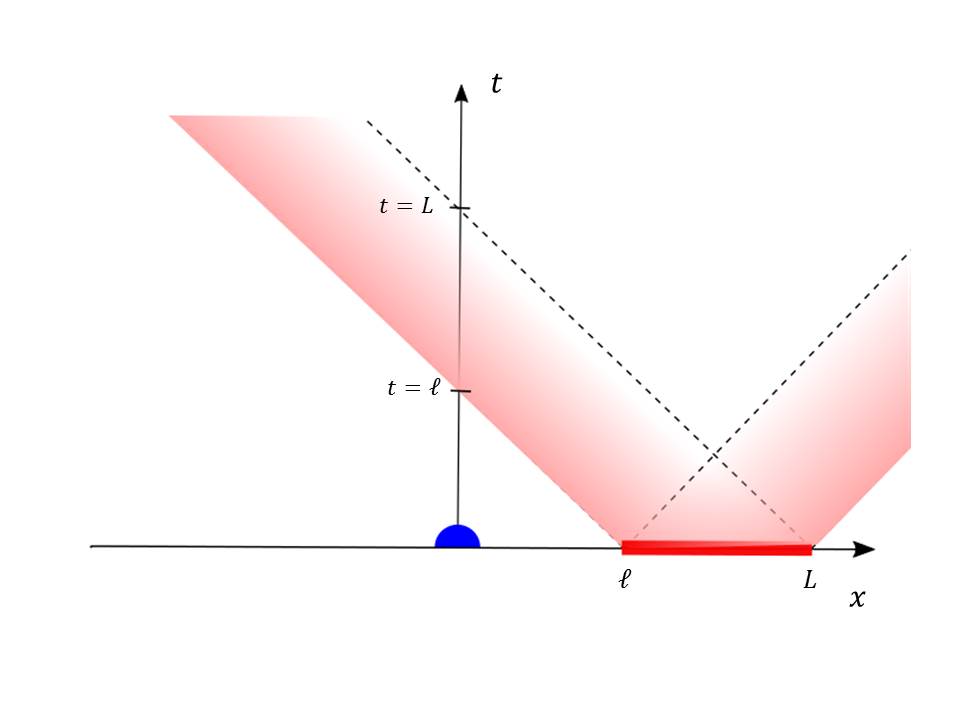}
  \caption{An operator is inserted at the origin. We consider its time evolution and measure the entanglement entropy for a subsystem $A= (\ell,L)$. The red area is the light-cone region of $A$. At time $t=\ell$ the operator crosses the light-cone and then crosses it on its way out at $t=L$ }
  \label{eesetup}
\end{figure}

We consider the EE of the subsystem $A$ which we take to be an interval $\ell\leq x\leq L$ at zero time in the presence of an excitation created by the energy momentum $T$ acting on the vacuum $\ket{0}$. Notice that since the energy momentum tensor is holomorphic, the computations involve only one complex coordinate. The location of the excitation will be at $x=0$ and we wish to consider the real time evolution of the entanglement entropy (see figure \ref{eesetup}).

The density matrix corresponding to this setup is 
\eq{\rho(t)&=\mathcal{N}\cdot e^{-iHt}e^{-\epsilon H}T(0)\ket{0}\bra{0}T(0)e^{-\epsilon H}e^{iHt}\\
&=\mathcal{N}\cdot T(w_2)\ket{0}\bra{0}T(w_1)\quad,}
where the normalization $\mathcal{N}$ is determined by requiring $Tr\rho(t)=1$ and the coordinates are defined by
\eq{w_1=i(\epsilon-it),\quad w_2=-i(\epsilon+it)\quad.}
As in the previous section, we will treat the Euclidean time $\epsilon-it$ as a real number till the very end where we will analytically continue by taking $\epsilon$ to zero. We keep in mind that $\epsilon$ is much smaller than all other parameters that we will introduce.

\begin{figure}
  \centering
      \includegraphics[width=\textwidth]{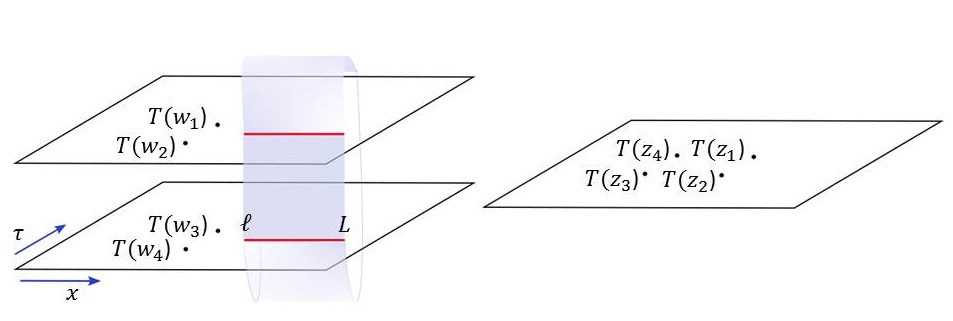}
  \caption{On the left is the two sheeted geometry which is used to calculate the $2$-nd R\'enyi entropy. The red line is the subsystem $A$ and the insertion of the energy momentum operator is smeared around $\tau=0$ in the time direction. On the right is the single sheet we encounter once performing the conformal transformation \eqref{ct}. In the case of the second Re\'nyi entropy, two points out of the four are reflections of the other two.}
  \label{2nd}
\end{figure}

We will calculate the difference between the entanglement entropy in the excited state to that in the vacuum state. 
The $n$-th R\'enyi entropy can be calculated using the Replica trick introduced in \cite{Calabrese:2004eu} which was generalized for excited states in \cite{Nozaki:2014hna}.
\eql{nRE}{\Delta S_A^{(n)}=\frac{1}{1-n}\left[\log\langle T(w_1)T(w_2)....T(w_{2n-1})T(w_{2n})\rangle_{\Sigma_n}-n\log\langle T(w_1)T(w_2)\rangle_{\Sigma_1}\right]\quad,}
where $\Sigma_n$ is an $n$-sheeted Riemann surface with two operators inserted on each sheet. These are located at $(w_{2k-1},w_{2k})$ on the $k$-th sheet for $k=1...n$ (see figure \ref{2nd}.)  The second term in this expression is just the two point function on a single sheet.
Naively we will expect the entanglement entropy to split into two regions: In the first, in which the excitation is outside the light-cone of the interval, we expect the EE to be the same as that of the vacuum state, i.e $\Delta S_A^{(n)}=0$. In the second region, in which the excitation is within the light-cone of the interval, we expect to get a non-zero $\Delta S_A^{(n)}$.

In order to calculate the correlation function over $\Sigma_n$, we apply the conformal transformation 
\eql{ct}{\frac{w-\ell}{w-L}=z^n\quad,}
which gives us $n$ coordinates $z_{2k-1}$ with $k=1...n$ corresponding to $w_1$ and $n$ coordinates $z_{2k}$ corresponding to $w_2$.
Using the transformed energy momentum tensor \eqref{T_trans}, the R\'enyi entropy \eqref{nRE} now takes a computable form
\eq{\Delta S_A^{(n)}=\frac{1}{1-n}\left[\log\langle\prod_{i=1}^{2k}\left(\frac{dw_i}{dz_i}\right)^{-2}\left[T(z_i)-\frac{c}{12}\{w_i,z_i\}\right]\rangle_{\Sigma_1}-n\log\langle T(w_1)T(w_2)\rangle_{\Sigma_1}\right]\quad.}
These energy momentum correlation functions can be computed by the recursion relation
 \eq{\langle T(z_1)T(z_2)...T(z_{2n})\rangle=&\sum_{j=2}^{2n}\frac{c/2}{(z_1-z_j)^4}\langle T(z_2)T(z_3)...T(z_{j-1})T(z_{j+1})...T(z_{2n})\rangle\\&+\left(\frac{2}{(z_1-z_j)^2}	+\frac{\partial_{z_j}}{z_1-z_j}\right)\langle T(z_2)T(z_3)...T(z_{2n})\rangle\quad.}
 Therefore, theoretically we can compute the R\'enyi entropy for any $n$.
 The second R\'enyi entropy involves the four point function with coordinates

\eq{&z_1=-z_3=\sqrt{\frac{\ell-t-i\epsilon}{L-t-i\epsilon}}\quad,\\
&z_2=-z_4=\sqrt{\frac{\ell-t+i\epsilon}{L-t+i\epsilon}}\quad,}
for which we can compute the cross ratio
\eq{z=\frac{z_{12}z_{34}}{z_{13}z_{24}}\rightarrow\frac{-z_{12}^2}{4z_1z_2}\quad.}
In terms of this cross ratio, the R\'enyi entropy is 
\begin{dmath}\Delta S_A^{(2)}=-\log\left(\frac{8}{c}z^2(1-z)^2(1-z+z^2)+(1-z)^4(1+z)^4+z^4
+8z^3(1-z)^3+cz^2(1-z)^2\left[(1-z)^2(1+z)^2+z^2\right]+\frac{c^2}{4}(1-z)^4z^4\right)\quad.
\end{dmath}
When calculating $z$ we see that it takes different values, depending on the time $t$. Assuming small $\epsilon$ this is $z=\frac{(L-\ell)^2\epsilon^2}{4(\ell-t)^2(L-t)^2}$ for $t<\ell$ and $t>L$, which is the regime in which the interval "doesn't know" about the excitation and $z=1-\frac{(L-\ell)^2\epsilon^2}{4(\ell-t)^2(L-t)^2}$ for $\ell<t<L$ which is the regime in which the excitation is within the light-cone of the interval. The existence of these two regimes is compatible with our naive expectation. But, surprisingly, these two values of $z$ turn out to give the same value for $\Delta S_A^{(2)}$. Specifically, when taking the limit $\epsilon\rightarrow 0$, we get
\eq{\Delta S_A^{(2)}\rightarrow 0\quad.}
This result has been reproduced also for $n=3$ (See Appendix \ref{n3}) and in the next section we will show that it is true also for the EE.
It is important to note that this result holds only in the case that the distances $\ell-t$ and $L-t$ are non-zero. In other words, the difference in EE is zero only away from the light-cone. Near the light-cone we get a diverging behaviour similar to the one we encountered for the energy flux where the result depends on the choice of our regulator $\epsilon$. In this case of the $2$-nd R\'enyi entropy this is the light-cone divergence of the four point function \cite{Hartman:2015lfa,Komargodski:2012ek,Fitzpatrick:2012yx,Li:2015itl}. 

\section{R\'enyi Entropy for Descendant Operators}\label{EEdes}
Following the results of the previous section, we attempt to generalize this behaviour to any descendant operator.
Far away from the light-cone we can compute the R\'enyi and entanglement entropies in an elegant way introduced in \cite{Sheikh-Jabbari:2016znt,Mandal:2014wfa,Rashkov:2016xnf,Beach:2016ocq}. Just as we did for the energy flux, a descendant operator can be constructed by a conformal transformation operator $U_f$ acting on a primary field $|\Psi\rangle$ in the cyclic orbifold theory $CFT^n/Z_n$. If we consider an excitation at the origin and an interval $A=(z_1,z_2)$ then the $n$-th R\'enyi entropy for an excited state takes the form  
\eq{\exp\left((1-n)S_{f\; A}^{(n)}(\Psi; z_1,z_2)\right)=\frac{\langle\Psi|U_f^\dagger\sigma_n(z_1)U_fU^\dagger_f\sigma_{-n}(z_2)U_f|\Psi\rangle_{\mathbb{C}}}{\langle\Psi|\Psi\rangle^n}\quad.}
Since $\sigma_n$ is a primary field of dimensions $\Delta_n=\frac{c}{24}\left(1-\frac{1}{n^2}\right)$ it transforms under the unitary operator as
\eq{U^\dagger_f\sigma_n(z) U_f=\left(\partial f(z)\right)^{\Delta_n}\left(\partial \bar{f}(\zb)\right)^{\Delta_n}\sigma(f(z))\quad,} 
therefore
\eql{onei}{S_{f\;A}^{(n)}(\Psi; z_1,z_2)&=S_{\; A}^{(n)}(\Psi; f(z_1),f(z_2))+\frac{1}{1-n}\frac{c}{12}\left(n-\frac{1}{n}\right)\log|\partial f(z_1)\partial f(z_2)|\quad.}
The first term is yet unknown for a generic primary $\Psi$ but if we consider descendants of the unity operator, then we can replace it with the entanglement entropy in the vacuum state with interval end points at $(f(z_1),f(z_2))$ which is given by \cite{Calabrese:2004eu}. This results in
\eq{S_{f\;A}^{(n)}(z_1,z_2)&=\frac{1}{1-n}\left(-\frac{c}{6}\left(n-\frac{1}{n}\right)\log\frac{|f(z_1)-f(z_2)|}{a}+\frac{c}{12}\left(n-\frac{1}{n}\right)\log|\partial f(z_1)\partial f(z_2)|\right)\\
&=-\frac{c}{12}\frac{1}{1-n}\left(n-\frac{1}{n}\right)\log\frac{|f(z_1)-f(z_2)|^2}{|a^2\partial f(z_1)\partial f(z_2)|}\quad.}
The EE is given by taking the limit $n\rightarrow 1$
\eql{Sof1}{S_{f\; A}(z_1,z_2)=\frac{c}{6}\log\frac{|f(z_1)-f(z_2)|^2}{|a^2\partial f(z_1)\partial f(z_2)|}\quad.}
This is the well-known result from \cite{Holzhey:1994we}.
If we wish to connect to our calculations in the previous section, with the excited state created by the energy momentum tensor, we need to take the appropriate conformal map $f(z)=\frac{1}{z},\quad \bar{f}(\zb)=\zb$. In this case 
\eql{S_vac}{S_{f\; A}(z_1,z_2)=\frac{c}{3}\log\frac{(z_1-z_2)}{a}\quad,} which is the same as the entanglement entropy in the vacuum, i.e $\Delta S_A^{(n)}=0$ for all $n$, in accordance with our results. Actually, for any $SL(2,\mathbb{C})$ transformation we will get the same result since the vacuum is invariant under $SL(2,\mathbb{C})$. We stress that this result is appropriate only away from the light-cone. In order to do a full analysis including the light-cone behaviour, we must use the regularization, by smearing the operator on the complex plane as we did in section \ref{EEEM}. 

The result we get for one interval \eqref{onei} can be extended to multiple disjoint intervals following \cite{Calabrese:2009ez, Calabrese:2010he}. For a subsystem $A=[z_1,z_2]\cup[z_3,z_4]\cup...\cup[z_{2N-1},z_{2N}]$ with $z_1<z_2<...<z_{2N}$ we can calculate the EE for a descendant of a primary $|\Psi\rangle$ 
\eq{\exp&\left((1-n)S_{f\; A}^{(n)}\right)\\&=\frac{\langle\psi| U_f^\dagger\sigma_n(z_1)U_fU^\dagger_f\sigma_{-n}(z_2)U_f...U_f^\dagger\sigma_n(z_{2N-1})U_fU^\dagger_f\sigma_{-n}(z_{2N})U_f|\Psi\rangle_{\mathbb{C}}}{\langle\Psi|\Psi\rangle^n}\\
&=\frac{\displaystyle\prod_{i=1}^{2N}|\partial f(z_i)|^{2\Delta_n}\langle\psi|\sigma_n\left(f(z_1)\right)\sigma_{-n}\left(f(z_2)\right)...\sigma_n\left(f(z_{2N-1})\right)\sigma_{-n}\left(f(z_{2N})\right)|\Psi\rangle_{\mathbb{C}}}{\langle\Psi|\Psi\rangle^n}\quad.}
We learn that also for multiple intervals the EE splits into two parts as following:
\eql{multiEE}{ S_{f\; A}(\Psi;z_1,...z_{2N})=S_{A}(\Psi;f(z_1),...f(z_{2N}))-\frac{c}{6} \log\displaystyle\prod_{i=1}^{2N}|\partial f(z_i)|}
This is a general formula for the EE for descendent operators in the case of multiple disjoint intervals, away from the light-cone. The first part is given by the EE of the primary field where the insertion points of the intervals are shifted by the conformal map $f$. The second  part is the addition to the EE coming from the descendant field alone.
As in the case of one interval, we expect singularities around the points where the excitation crosses the light-cones of the intervals but we will not prove this here.

\section{Flux to EE}\label{fluxEE}
As we saw in the previous sections, the energy flux and the EE both exhibit a singularity on the light-cone. Moreover, we recall \cite{He:2014mwa,Nozaki:2014hna} that for an entangled state $\mo_1=e^{\frac{i}{2}\phi}$ where $\phi$ is a (non-compact) free massless scalar, the entanglement entropy difference vanishes in all regions $\Delta S_A=0$. On the other hand, for the state created by $\mo_2=\onov{\sqrt{2}}(e^{\frac{i}{2}\phi}+e^{-\frac{i}{2}\phi})$, the entanglement entropy difference is 
\[\Delta S_A=\left\{\begin{array}{lr}
0&(0<t<\ell\quad or\quad t>L)\quad,\\
\log 2& (\ell<t<L)\quad.
 \end{array}\right.\]
These two operators have the same conformal dimensions and therefore, by \eqref{flux_p} they have the same energy flux. Away from the light-cone this flux is zero but if we consider descendants of these fields, generated by some function $f$, they will still have the same energy flux and it can be non-zero away from the light-cone. This suggests that the energy flux is related to the EE by derivatives. 
In the large $c$ limit, the EE has separated holomorphic and anti-holomorphic parts therefore it will be convenient to define the two parts of the EE in the following way
\eq{S^{tot}_{f\; A}&(\Psi; z_1,z_2)=S_{f\; A}(\Psi; z_1,z_2)+\bar S_{\fb\; A}(\Psi; \zb_1,\zb_2)\\&=S_A(\Psi; f(z_1),f(z_2))+\Sb(\Psi;\fb(\zb_1),\fb(\zb_2))-\frac{c}{12}\log(\partial f(z_1)\partial f(z_2))-\frac{c}{12}\log(\bar\partial\fb(\zb_1)\bar\partial\fb(\zb_2))\quad.}
Without loss of generality, we will consider only the holomorphic parts for which we suggest the ansatz
\eql{ans}{\langle \mo_f T(z)\mo_f\rangle=-\frac{\partial^2 S_{f\; A}(z,z')}{\partial z^2}-\frac{6}{c}\left(\frac{\partial S_{f\; A}(z,z')}{\partial z}\right)^2\quad,}
with $\mo_f$ being the operator creating an excited state by a descendant field: $\mo$ is the primary and $f$ is the function associated with the descendant of $\mo$. Since this is a second order differential equation, we need to specify two conditions as we will see below. Intuitively this is obvious since the energy flux is determined by one parameter: the distance between the calorimeter and the source, while the EE is determined by two: The distance from the source and the length of the subsystem $A$. This relation has been stated in \cite{Bianchi:2014qua} in the context of black holes and it holds only away form the light-cone.
 We will consider some cases for which we know the EE for which \eqref{ans} is true. 
\begin{itemize}
\item {\bf The vacuum state}\\
In the vacuum the energy flux is zero, therefore we get a homogeneous second order differential equation which can be easily solved. One solution is $S=constant$ which we will disregard since it doesn't describe a physical system. The other solution is 
\eq{S= \frac{c}{6}\log(z+B)+D\quad,}
 	where $B$ and $D$ are constants that we can determine by appropriate conditions. For $z=z'$ i.e when the interval is zero we will demand $S(z=z')=-\infty$. With the other condition we can set $D$ so that we get the well-known result (for the holomorphic part)
 	\eq{S(z,z')=\frac{c}{6}\log\frac{(z-z')}{\sqrt{a}}\quad.}
\item {\bf Descendants of the unit operator}\\
In this case, the energy flux is just the Schwarzian derivative \eq{\langle \mathbbm{1}_f|(T(z)|\mathbbm{1}_f\rangle=\frac{c}{12}\{f(z),z\}\quad,}
therefore \eqref{ans} becomes a non-homogeneous differential functional equation which we can write as
\eq{-\frac{\partial^2 S}{\partial z^2}-\frac{6}{c}\left(\frac{\partial S}{\partial z}\right)^2&=-\frac{\partial^2 S}{\partial f^2}(f')^2-\frac{\partial S}{\partial f}f''-\frac{\partial^2S}{\partial f^{'2}}(f'')^2-\frac{\partial S}{\partial f'}f'''-\frac{6}{c}\left(\frac{\partial S}{\partial f}f'+\frac{\partial S}{\partial f'}f''\right)^2\\&=\frac{c}{12}\{f(z),z\}\quad.}
The EE we found in \eqref{Sof1} for the descendants of the unit operator solves this equation in agreement with our ansatz.
\end{itemize}
It is tempting to generalize this relation to all excited states but we do not yet have a general form for the EE for excited states by primary operators in CFTs.

Nevertheless, we can use holography to extend this relation to all excited states related to Ba\~nados geometries with constant $L_\pm$, mentioned earlier in \eqref{banados}. These correspond to the following geometries \cite{Banados:1992wn,Banados:1992gq,Deser:1983nh}: $L_+=L_-=-1/4$ correspond to $AdS_3$ in global coordinates, $-1/4<L_\pm<0$ correspond to conical defects, $L_+=L_-=0$ correspond to massless BTZ black holes and positive $L_\pm=\mathfrak{T}_\pm^2$ correspond to generic BTZ black holes. 
The EE in these cases has been computed \cite{Sheikh-Jabbari:2016znt} and results in \footnote{For negative $L_\pm$ we take imaginary $\mathfrak{T}_\pm$.}
\eq{S=\frac{c}{6}\log \left(\frac{\sinh(\mathfrak{T}_+R_+)}{\mathfrak{T}_+\epsilon}\frac{\sinh(\mathfrak{T}_-R_-)}{\mathfrak{T}_-\epsilon}\right)\quad,}
where $R_\pm=\min(\Delta x^\pm,2\pi-\Delta x^\pm)$ and $\Delta x^\pm=|x_1^\pm-x_2^\pm|$.  Focusing on the holomorphic part, one may convince himself that if we plug this result into 
\eqref{ans} then we get $\langle\mo_f T(x_1^+)\mo_f\rangle=\frac{c}{6}\mathfrak{T}_+^2$ in agreement with \eqref{banadosT}. We can now state that our ansatz is correct for CFTs dual to geometries with constant representatives, for primary states and their descendants.

 
Moreover, this kind of relation for descendants of the vacuum can be generalized to multi-intervals using \eqref{multiEE} and \eqref{Nflux}. A general form of the EE in the vacuum for multiple intervals is not known but for some cases it has been computed, see 
\cite{Calabrese:2009ez,Calabrese:2010he} and references therein. The EE for two disjoint intervals $A=[z_1,z_2]\cup[z_3,z_4]$ in the large $c$ limit has been calculated both in the CFT theory and holographically  \cite{Headrick:2010zt,Hartman:2013mia} and takes the form 
\eql{SRT}{S_A=\text{min}\left(S_{12}+S_{34}, S_{14}+S_{23}\right)\quad,}
where $S_{ij}$ is the EE for a single interval with endpoints $z_i$ and $z_j$. 
Our ansatz for the holomorphic part for two intervals in this limit will be 
\eql{ans2}{\langle\mo_f T(z_1)T(z_3)\mo_f\rangle=\left(-\frac{\partial^2 S_A}{\partial z_1^2}-\frac{6}{c}\left(\frac{\partial S_A}{\partial z_1}\right)^2\right)\left(-\frac{\partial^2 S_A}{\partial z_3^2}-\frac{6}{c}\left(\frac{\partial S_A}{\partial z_3}\right)^2\right)\quad.}
We follow the same procedure as for the one interval for descendants of the operators we considered for the one interval. The non-homogeneous part is given by the energy flux \eqref{2flux} which in the large $c$ limit is
 \eq{\langle\mo_f T(z_1)T(z_3)\mo_f\rangle\sim\left(\frac{c}{12}\right)^2\{f(z_1),z_1\}\{f(z_3),z_3\}\quad.}
This equation has four initial conditions to specify. To make contact with the known result \eqref{SRT} we can choose 
\eq{S(f(z_1)= f(z_2), f(z_3)= f(z_4))\rightarrow-\infty\quad
\text{or}\quad
S(f(z_1)= f(z_4),f(z_3)= f(z_2))\rightarrow-\infty\quad,}
together with the other two initial conditions that we chose for the one interval case. We get
\eq{S=&\frac{c}{6}\log\left((f(z_1)-f(z_2))(f(z_3)-f(z_4))\right))-\frac{c}{12}\log(f'(z_1)f'(z_2)f'(z_3)f'(z_4))\quad,\\
&\text{or}\\
S=&\frac{c}{6}\log\left((f(z_1)-f(z_4))(f(z_3)-f(z_2))\right))-\frac{c}{12}\log(f'(z_1)f'(z_2)f'(z_3)f'(z_4))\quad,}
in accordance with \eqref{multiEE} and \eqref{SRT}.

This can be generalized further to any amount of intervals, in the large $c$ limit. In this limit the EE decouples into a sum of single-interval entanglements and the energy flux is a product of Schwarzian derivatives. Therefore it is natural to generalize \eqref{ans2} to 
\eq{\langle\mo_f T(z_1)T(z_3)...T(z_{2N-1})\mo_f\rangle=\prod_{i=1}^{2N-1}\left(-\frac{\partial^2 S_A}{\partial z_i^2}-\frac{6}{c}\left(\frac{\partial S_A}{\partial z_i}\right)^2\right)\quad,}
where the product is over odd indices alone.
  Choosing the right initial conditions, we recover \eqref{multiEE} with the first term given by Hartman's formula \cite{Hartman:2013mia} (for the holomorphic part)
\eq{S=\text{min}\frac{c}{6}\sum_{(i,j)}\log\left(\frac{z_i-z_j}{\epsilon}\right)\quad.}  The sum is over pairs $(i,j)$ dictated by the OPE channel as described in the original paper. 

To summarize, in this section we suggested a relation between the energy flux and the EE in the vacuum and for descendants of the vacuum for one, two and multiple intervals. The same relation holds for rational CFTs and their descendants in a setup with one interval.


\section{Summary and Outlook}
In this work we have studied properties of EE for excited states in 2D CFTs. We summarize our main results:
\begin{itemize}
\item
We have pointed out that for one interval in the presence of an excited state the EE has a singularity on the light-cone.
\item
A formula expressing the EE of a descendant field in terms of the EE of its primary has been established for multiple intervals, away from the light-cone \eqref{multiEE}.
We have noticed that the EE does not enjoy contributions from excited states which are descendants of the unit operator.
\item 
We suggest a relation \eqref{ans} between the EE and the energy flux created by an excited state. This relation holds in the following cases: The vacuum state, i.e  no excitation; an excitation created by descendants of the unit operator and in the large $c$ limit, excitations of primaries related holographically to Ba\~nados geometries with constant representatives and their descendants. 
\item In the case of the vacuum and descendants of the unit operator we were able to generalize the relation to multiple intervals relying on the large $c$ limit.

\end{itemize}

The relationship between EE and energy flux has been investigated in the past resulting in a first-law-like relation \cite{Bhattacharya:2012mi}. It would be interesting to connect these two different approaches to one coherent picture. Furthermore, one might try to generalize the relation to generic excited states in 2D conformal field theories.

\section*{Acknowledgement}
We are grateful to Z. Komargodski, M. Mezei, A. Sever, J. Sim\'on and G. Torrents for fruitful discussions. T.V. is supported by the ERC STG grant 335182.
\appendix
\section{$n=3$}\label{n3}

\begin{figure}
  \centering
      \includegraphics[width=0.5\textwidth]{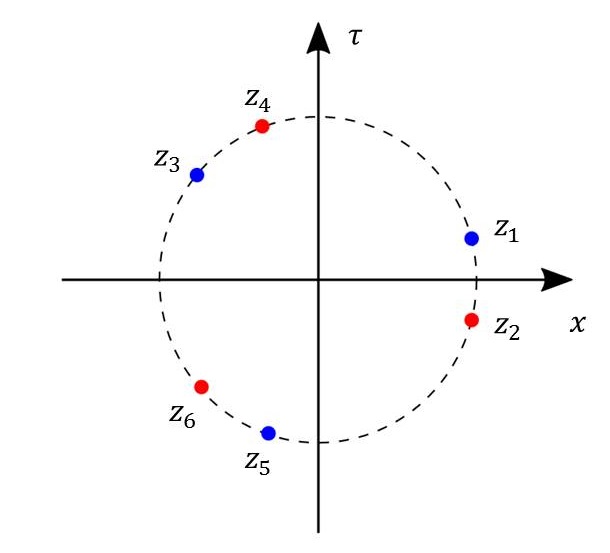}
  \caption{For the 3rd R\'enyi entropy, the three sheets are mapped to one sheet with the six points as depicted. They are two sets of 3rd roots of unity.}
  \label{3rd}
\end{figure}

The third R\'enyi entropy is given by a six-point function over the three-sheeted manifold. In this case we use the mapping
\eq{\frac{w-\ell}{w-L}=z^3\quad,} which yields the six coordinates of the operator insertions (see figure \ref{3rd})
\eq{&z_1=\left(\frac{\ell-t-i\epsilon}{L-t-i\epsilon}\right)^{1/3},\quad z_2=\left(\frac{\ell-t+i\epsilon}{L-t+i\epsilon}\right)^{1/3}\quad,\\&z_3=e^{\frac{2\pi i }{3}}\left(\frac{\ell-t-i\epsilon}{L-t-i\epsilon}\right)^{1/3},\quad z_4=e^{\frac{2\pi i }{3}}\left(\frac{\ell-t+i\epsilon}{L-t+i\epsilon}\right)^{1/3}\quad,\\
&z_5=e^{\frac{4\pi i }{3}}\left(\frac{\ell-t-i\epsilon}{L-t-i\epsilon}\right)^{1/3},\quad z_6=e^{\frac{4\pi i }{3}}\left(\frac{\ell-t+i\epsilon}{L-t+i\epsilon}\right)^{1/3}\quad.\\
}

The R\'enyi entropy is given by $\Delta S^{(3)}_A=-\frac{1}{2}\log Tr[\rho_A^3]$ with
\eq{Tr[\rho_A^3]=&\frac{\langle T(w_1)T(w_2)T(w_3)T(w_4)T(w_5)T(w_6)\rangle_{\Sigma_3}}{\left(\langle T(w_1)T(w_2)\rangle_{\Sigma_1}\right)^3}\\&=\prod_{i=1}^6\left(\frac{dw_i}{dz_i}\right)^{-2}\left[T(z_i)-\frac{c}{12}\{w_i,z_i\}\right]\left(\frac{c/2}{\left(\frac{Lz_1^3-l}{z_1^3-1}-\frac{Lz_2^3-l}{z_2^3-1}\right)^4}\right)^{-3}\quad.}
We see that we get contribution from $0,2,3,4,5$ and $6$ point functions. When expanding for small $\epsilon$, assuming that $\epsilon<<|\ell-t|,|L-t|$, it turns out that all of them vanish up till order $\mathcal{O}(\epsilon^4)$ all but the $6$-point function that contributes $1$. Therefore
\eq{\Delta S^{(3)}_A=0\quad.}
But when taking $|\ell-t|\quad\text{or}\quad |L-t|<<\epsilon$  then we reach a diverging behaviour, similar to the one we found for the second R\'enyi entropy.
\bibliographystyle{unsrt}
\bibliography{bibEE}
\end{document}